\documentclass[english,twocolumn,showpacs,preprintnumbers,amsmath,amssymb,aps]{revtex4-1}
\usepackage[T1]{fontenc}
\usepackage[latin9]{inputenc}
\usepackage{color}
\usepackage{array}
\usepackage{multirow}
\usepackage{amssymb}
\usepackage{graphicx}

\makeatletter

\providecommand{\tabularnewline}{\\}

 \@ifundefined{textcolor}{}
 {%
   \definecolor{BLACK}{gray}{0}
   \definecolor{WHITE}{gray}{1}
   \definecolor{RED}{rgb}{1,0,0}
   \definecolor{GREEN}{rgb}{0,1,0}
   \definecolor{BLUE}{rgb}{0,0,1}
   \definecolor{CYAN}{cmyk}{1,0,0,0}
   \definecolor{MAGENTA}{cmyk}{0,1,0,0}
   \definecolor{YELLOW}{cmyk}{0,0,1,0}
 }

\makeatother

\usepackage{palatino}
\usepackage{babel}
\begin{document}

\title{On the empirical dipole polarizability of He from spectroscopy of
HeH$^{+}$}

\author{Nikesh S. Dattani}

\email{nike.dattani@hertford.ox.ac.uk}

\selectlanguage{english}%

\affiliation{Physical and Theoretical Chemistry Laboratory, Department of Chemistry,
Oxford University, OX1 3QZ, Oxford, UK, }

\affiliation{Quantum Chemistry Laboratory, Department of Chemistry, Kyoto University,
606-8502, Kyoto, Japan,}

\author{Mariusz Puchalski}

\email{mariusz.puchalski@fuw.edu.pl}

\selectlanguage{english}%

\affiliation{Faculty of Chemistry, Adam Mickiewicz University, Umultowska 89b,
61-614 Pozna\'{n}, Poland. }
\begin{abstract}
Using a long-range polarization potential for HeH$^{+}$, we can obtain
an empirical value for the Boltzmann constant $k_{B}$ with an order
of magnitude greater precision than the previous best experimental
value based on the dipole polarizability of $^{4}$He. We find that
relativistic and QED effects of order $\alpha_{{\rm FS}}^{3}$ in
the fine structure constant are crucial in the quadrupole polarizability
in order to fit the dipole polarizbility with this precision using
the polarization potential. By calculating finite-mass corrections
for $^{3}$He, we also present theoretical values for the dipole and
quadrupole polarizabilities of $^{3}$He with 9 and 7 digits of precision
respectively.
\end{abstract}

\pacs{06.20.Jr, 31.30.jh, 31.50.Bc , 95.30.Ky }

\maketitle
In November 2014, the SI system of scientific units is expected to
undergo the biggest change in its 64 year history. Currently, temperature
is defined based on the triple point of water, which is rather arbitrary,
and due to the precision at which this property is known, the current
definition of temperature is unsatisfactory for temperatures below
20 K and above 1300 K. 

In the new system, the Boltzmann constant ($k_{B})$ will be defined
to be exact, and the triple point of water is derived from $k_{B}$,
rather than the other way around. Paramount to this new SI system,
is the need for a precise and reliable value of $k_{B}$. Currently
the relative uncertainty in $k_{B}$ is 9.1$\times10^{-7}$ which
is more than an order of magnitude larger than other fundamental constants
(for example, it is 4.4$\times10^{-8}$ for the Avogadro constant
$N_{{\rm A}}$, and 2.2$\times10^{-8}$ for the elementary charge
$e$). A much more reliable value of $k_{B}$ can be determined from
a high-precision measurement of the static dipole polarizability $\alpha_{1}$
of a substance: 

\begin{equation}
k_{B}=\alpha_{1}\frac{N_{{\rm A}}^{2}(\epsilon_{r}+2)}{3p\epsilon_{0}(\epsilon_{r}-1)},\label{eq:kB}
\end{equation}
where the vacuum permittivity $\epsilon_{0}$ is defined exactly,
the pressure $p$ at which the experiment is conducted can be reliably
held fixed, and the refractive index of the substance can be measured
extremely accurately. Gaseous helium has received enormous attention
as a candidate substance for the experiment which re-defines $k_{B}$,
because $\alpha_{1}$ is known theoretically to at least 8 digits
\cite{Lach2004} and experimentally to at least 6 \cite{Schmidt2007}. 

In 2009 the resonance dipole-dipole interaction $C_{3}$ of Li was
empirically determined with a higher precision than any previously
determined oscillator strength for any system, by an order of magnitude
\cite{LeRoy2009,Tang2011}. This was made possible by a fit of 17~477
high-precision spectroscopic measurements to a potential with a correct
theoretical model for the long-range potential built into it: the
longest-range term of the model was $-C_{3}/r^{3}$, so the fitted
value of $C_{3}$ was based on these 17~477 high-precision measurements.
The case for He is similar, except instead of $C_{3}$ we are interested
in $\alpha_{1}$ which is related to the longest-range term $-\frac{1}{2}\alpha_{1}/r^{4}$
of the polarization potential of HeH$^{+}$. The more accurate the
model for the polarization potential, the better the determination
of $\alpha_{1}$ by the fit.

The theory for the polarization potential between a neutral atom and
a far away charge is related to the theory for an ionic core interacting
with a far away electron in a Rydberg atom, and has a rich history
dating back to at least 1933 \cite{Mayer1933}. At that time the adiabatic
potential from the electric multipoles for systems with a zero angular
momentum core was derived up to $1/r^{6}$. Non-adiabatic corrections
up to $1/r^{-6}$ were derived in 1968 independently by Kleinman
\emph{et al.} and Dalgarno \emph{et al.} \cite{Dalgarno1968}\emph{,
}and the full electric multipole potential including both adiabatic
and non-adiabatic contributions up to $1/r^{8}$ , for systems where
the core has no angular momentum, was first derived in 1982 by Drachman
\cite{Drachman1982}. 

\begin{figure*}
\caption{Empirical internuclear potential for the ground electronic state of
$^{4}$HeH$^{+}$ and the theoretical long-range polarization potential
using varying numbers of terms in Eq. \ref{eq:polarizationPotential}.
All long-range coefficients are calculated from Eq. \ref{eq:polarizationPotential}
using the most precise current values of the polarizabilities, presented
in Table \ref{tab:polarizabilitiesFiniteMass}.\label{fig:potentials}}

\includegraphics[width=1\textwidth]{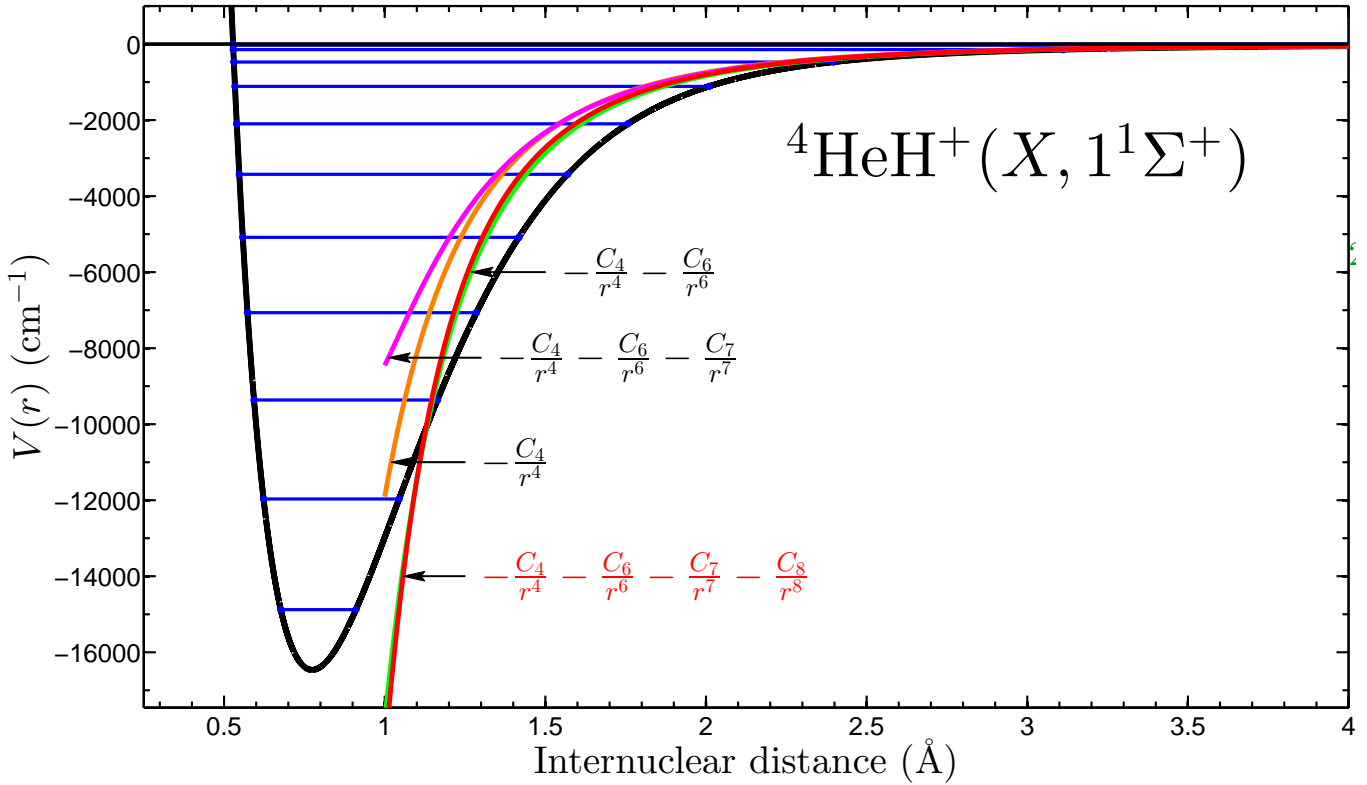}
\end{figure*}

In 2012 Woods and Lundeen \cite{Woods2012} have derived corrections
to this potential for systems where the core has non-zero angular
momentum. Drachman's potential is only the result of the first order
in the tensor expansion of Woods and Lundeen, but the latter derived
adiabatic and non-adiabatic terms up to $1/r^{8}$, and up to 5th
order in the tensor expansion. Woods and Lundeen also presented energy
corrections due to relativity and due to various magnetic interactions.
It turns out that all of these post-Drachman terms are zero for all
known isotopes of He($^{1}S$) with a half-life of $>1$ ms, due to
the nuclear spin always being $<\frac{3}{2}$, the lack of additional
angular momentum, and the lack of an electric charge.

The rotationless ($J=0$) polarization potential for any isotopologue
of HeH$^{+}$ including adiabatic and non-adiabatic terms up to $1/r^{8}$,
due to the electric multipoles of He is therefore \cite{Drachman1982,Tang2010b}:

\begin{widetext}

\begin{eqnarray}
V_{{\rm pol}}(r) & = & -\frac{C_{4}}{r^{4}}-\frac{C_{6}}{r^{6}}-\frac{C_{7}}{r^{7}}-\frac{C_{8}}{r^{8}}-\cdots\\
 & = & -\frac{\alpha_{1}}{2r^{4}}-\frac{\alpha_{2}-\frac{3\beta_{1}}{\mathfrak{M}}}{2r^{6}}-\frac{\alpha_{112}}{2r^{7}}-\frac{\alpha_{3}-\frac{\beta_{2}}{\mathfrak{M}}-\alpha_{1}\alpha_{2}+\alpha_{1111}+\frac{72\gamma_{1}}{\mathfrak{M}}}{2r^{8}}-\cdots.\label{eq:polarizationPotential}
\end{eqnarray}

\end{widetext}

The reduced mass $\mathfrak{M}\equiv\frac{M_{{\rm He}}M_{{\rm H^{+}}}}{M_{{\rm He}}+M_{{\rm H^{+}}}}$,
which was first introduced for the $\beta_{1}$ term in \cite{Dalgarno1968},
significantly reduces the effect of the non-adiabatic terms (represented
by $\beta_{i}$ and $\gamma_{i}$). All other symbols are described
in the caption to Table \ref{tab:PolarizabilitiesInfiniteMass} and
formulas to calculate them are presented in \cite{Bhatia1994}. 

For large $r$, the potentials for the isotopologues of HeH$^{+}$
are dominated by the above polarization function. Therefore spectroscopic
measurements involving the long-range energy levels of HeH$^{+}$
can be fit to the eigenenergies of the Hamiltonian containing the
above energy functions with $\alpha_{1}$ treated as a free parameter.
No matter how precisely these spectroscopic measurements are made,
the $\alpha_{1}$ will be determined incorrectly if the models for
the polarization functions are not correct. Since there are so many
terms appearing in the polarization functions, it is difficult to
intuit which effects are necessary to include in each term in order
to determine $\alpha_{1}$ with a greater degree of confidence than
the current best empirical value, which was determined by a different
technique in 2007 \cite{Schmidt2007}.

Determining precisely which effects are necessary to include in the
polarization functions for this purpose is needed for two reasons:
\begin{enumerate}
\item It will provide an indication of which effects need to be included
in the calculation of a polarization model when undertaking similar
fitting endeavors for other systems.
\item Fitting model potentials to spectroscopic data becomes more difficult
as more terms are included in the long-range potential \cite{Dattani2011,LeRoy2009},
therefore it is always desired to have an \emph{a priori} indication
of how many terms need to be included.
\end{enumerate}
\begin{table*}
\begin{raggedright}
\caption{Theoretical polarizabilities from the 0th ($\alpha_{i}$) 1st ($\beta_{i}$)
and 2nd ($\gamma_{i}$) orders of the adiabatic expansion, various
orders ($2^{i}$) of the multipole expansion, and up to fourth order
perturbation theory. $\alpha_{112}$ and $\alpha_{1111}$ come from
third and fourth order perturbation theory respectively. All polarizabilities
are in Hartree atomic units. We used the NIST nuclear masses $M(^{4}{\rm He})=4.001506179125(62)$
a.u., $M(^{3}\mbox{He})=3.0149322468(25)$ a.u.\label{tab:PolarizabilitiesInfiniteMass}}

\par\end{raggedright}

\raggedright{}{\scriptsize }%
\begin{tabular*}{1\textwidth}{@{\extracolsep{\fill}}cclllllc}
\noalign{\vskip2mm}
 &  &  & \multicolumn{3}{c}{corrections} &  & \tabularnewline
\cline{4-6} 
\noalign{\vskip2mm}
 &  & {\scriptsize non-relativistic} & {\scriptsize $\alpha_{{\rm FS}}^{2}$ (relativistic)} & {\scriptsize $\alpha_{{\rm FS}}^{3}$ (QED)} & {\scriptsize $\alpha_{{\rm FS}}^{4}$ (QED)} & \multicolumn{2}{c}{\textbf{\scriptsize total (infinite mass)}}\tabularnewline[2mm]
\hline 
\hline 
\noalign{\vskip2mm}
\textbf{\scriptsize adiabatic} &  &  &  &  &  &  & \tabularnewline
{\scriptsize dipole} & {\scriptsize $\alpha_{1}$} & {\scriptsize $1.383192174455(1)$} & {\scriptsize $-8.035(2)\times10^{-5}$} & {\scriptsize $3.0666(1)\times10^{-5}$} & {\scriptsize $4.9(23)\times10^{-7}$} & \textbf{\textcolor{blue}{\scriptsize $\mathbf{1.38314230(23)}$}} & \textcolor{black}{\scriptsize \cite{Lach2004} }\tabularnewline
{\scriptsize quadrupole} & {\scriptsize $\alpha_{2}$} & {\scriptsize $2.4450831045(2)$} & {\scriptsize $-1.750786(2)\times10^{-4}$} & {\scriptsize $7.5(6)\times10^{-5}$} & {\scriptsize ~~~~-} & \textbf{\textcolor{blue}{\scriptsize $\mathbf{2.446858(6)}$}} & \textbf{\textcolor{black}{\scriptsize {[}this work{]}}}\tabularnewline
{\scriptsize octupole} & {\scriptsize $\alpha_{3}$} & {\scriptsize $10.6203286(2)$} & {\scriptsize ~~~~~~~~~~-} & {\scriptsize ~~~~~~-} & {\scriptsize ~~~~-} & \textcolor{blue}{\scriptsize $\mathbf{10.6203286(2)}$} & \textcolor{black}{\scriptsize \cite{Yan1996}}\tabularnewline
{\scriptsize hexadecapole} & {\scriptsize $\alpha_{4}$} & {\scriptsize $86.905$} & {\scriptsize ~~~~~~~~~~-} & {\scriptsize ~~~~~~-} & {\scriptsize ~~~~-} & \textcolor{blue}{\scriptsize $\mathbf{86.905}$} & \textcolor{black}{\scriptsize \cite{Kar2010}}\tabularnewline
{\scriptsize dip.-dip.-quad. } & {\scriptsize $\alpha_{112}$} & {\scriptsize $7.3267069796(16)$} & {\scriptsize ~~~~~~~~~~-} & {\scriptsize ~~~~~~-} & {\scriptsize ~~~~-} & \textbf{\textcolor{blue}{\scriptsize $\mathbf{7.3267069796(16)}$}} & \textbf{\textcolor{black}{\scriptsize {[}this work{]}}}\tabularnewline
{\scriptsize dip.-dip.-dip.-dip.} & {\scriptsize $\alpha_{1111}$} & {\scriptsize $4.57064147(2)$} & {\scriptsize ~~~~~~~~~~-} & {\scriptsize ~~~~~~-} & {\scriptsize ~~~~-} & \multirow{1}{*}{\textbf{\textcolor{blue}{\scriptsize $\mathbf{4.57064147(2)}$}}} & \textbf{\textcolor{black}{\scriptsize {[}this work{]}}}\tabularnewline[2mm]
\hline 
\noalign{\vskip2mm}
\textbf{\scriptsize non-adiabatic } &  &  &  &  &  &  & \tabularnewline
{\scriptsize dipole} & {\scriptsize $\beta_{1}$} & {\scriptsize $0.707510144012(5)$} & {\scriptsize ~~~~~~~~~~-} & {\scriptsize ~~~~~~-} & {\scriptsize ~~~~-} & \textbf{\textcolor{blue}{\scriptsize $\mathbf{0.707510144012(5)}$}} & \textbf{\textcolor{black}{\scriptsize {[}this work{]}}}\tabularnewline
{\scriptsize quadrupole} & {\scriptsize $\beta_{2}$} & {\scriptsize $1.0360960379(3)$} & {\scriptsize ~~~~~~~~~~-} & {\scriptsize ~~~~~~-} & {\scriptsize ~~~~-} & \textbf{\textcolor{blue}{\scriptsize $\mathbf{1.0360960379(3)}$}} & \textbf{\textcolor{black}{\scriptsize {[}this work{]}}}\tabularnewline[2mm]
\hline 
\noalign{\vskip2mm}
{\scriptsize dipole} & {\scriptsize $\gamma_{1}$} & {\scriptsize $0.385534894714(9)$} & {\scriptsize ~~~~~~~~~~-} & {\scriptsize ~~~~~~-} & {\scriptsize ~~~~-} & \textbf{\textcolor{blue}{\scriptsize $\mathbf{0.385534894714(9)}$}} & \textbf{\textcolor{black}{\scriptsize {[}this work{]}}}\tabularnewline
{\scriptsize quadrupole} & {\scriptsize $\gamma_{2}$} & {\scriptsize $0.4676192175(4)$} & {\scriptsize ~~~~~~~~~~-} & {\scriptsize ~~~~~~-} & {\scriptsize ~~~~-} & \textbf{\textcolor{blue}{\scriptsize $\mathbf{0.4676192175(4)}$}} & \textbf{\textcolor{black}{\scriptsize {[}this work{]}}}\tabularnewline
\end{tabular*}
\end{table*}

\begin{table*}
\caption{The effect of finite mass corrections yields final polarizabilities
for $^{3}\mbox{He}$ and $^{4}$He. The $^{4}$He dipole polarizability
is from \cite{Lach2004}. All polarizabilities are
in Hartree atomic units. \label{tab:polarizabilitiesFiniteMass}}

\noindent {\scriptsize }%
\begin{tabular*}{1\textwidth}{@{\extracolsep{\fill}}lllllll}
\noalign{\vskip2mm}
 &  & \multicolumn{2}{c}{$^{3}$He} &  & \multicolumn{2}{c}{$^{4}$He}\tabularnewline
\cline{3-4} \cline{6-7} 
\noalign{\vskip2mm}
 & Infinite mass & Finite mass correction & Final &  & Finite mass correction & Final\tabularnewline[2mm]
\cline{1-4} \cline{6-7} 
\noalign{\vskip2mm}
$\alpha_{1}$ & \textbf{\textcolor{blue}{1.38314230(23)}} & $8.2000572130(2)\times10^{-4}$ & \textcolor{red}{$\mathbf{1.38396299(23)}$} &  & $6.1781195345(2)\times10^{-4}$ & \textbf{\textcolor{red}{$\mathbf{1.38376079(23)}$}}\tabularnewline
$\alpha_{2}$ & \textbf{\textcolor{blue}{$\mathbf{2.444983(6)}$}} & $2.488746306(2)\times10^{-3}$ & \textbf{\textcolor{red}{$\mathbf{2.447472(6)}$}} &  & $1.874986955(2)\times10^{-3}$ & \textbf{\textcolor{red}{$\mathbf{2.446858(6)}$}}\tabularnewline
$\alpha_{112}$ & \textcolor{blue}{$\mathbf{7.3267069796(16)}$} & $8.92473393(2)\times10^{-3}$ & \textcolor{red}{$\mathbf{7.335631713567(2)}$} &  & $6.72361201(2)\times10^{-3}$ & \textcolor{red}{$\mathbf{7.3334305917(16)}$}\tabularnewline[2mm]
\hline 
\noalign{\vskip2mm}
$\beta_{1}$ & \textbf{\textcolor{blue}{$\mathbf{0.707510144012(5)}$}} & $5.3013314612(4)\times10^{-4}$ & \textcolor{red}{$\mathbf{0.708040277158(5)}$} &  & $3.9940700445(3)\times10^{-4}$ & \textcolor{red}{$\mathbf{7.07909551017(5)}$}\tabularnewline
\end{tabular*}
\end{table*}

\begin{table*}
\caption{Comparison of fitted values of $\alpha_{1}$ when fitting to various
models. In these models, $\alpha_{1}$ is treated as a free parameter
for the fit, and the rest of the polarizabilities come from Table
\ref{tab:polarizabilitiesFiniteMass}. $\alpha_{2}^{{\rm (no\, QED)}}$
is $\alpha_{2}$ with the QED term proportional to $\alpha_{{\rm FS}}^{3}$
ignored. Numbers in parentheses represent uncertainties in the last
digit(s) of the quantity shown. The uncertainties in the fits are
based only on the convergence of the fit. All polarizabilities are
in Hartree atomic units. \label{tab:fittedDipolePolarizability}}

\begin{tabular*}{1\textwidth}{@{\extracolsep{\fill}}>{\raggedright}p{0.5\textwidth}ll}
\hline 
\noalign{\vskip2mm}
 & $^{3}$He & $^{4}$He\tabularnewline[2mm]
\hline 
\noalign{\vskip2mm}
Theory  & \textcolor{red}{$\mathbf{1.38396299(23)}$} & \textbf{\textcolor{red}{$\mathbf{1.38376079(23)}$}} \cite{Lach2004}\tabularnewline
Experiment  & ~~~~~~~~- & $1.383759(13)$ \cite{Schmidt2007}\tabularnewline[2mm]
\hline 
\noalign{\vskip2mm}
{\footnotesize $-\frac{\alpha_{1}}{2r^{4}}-\frac{\alpha_{2}-\frac{3\beta_{1}}{\mathfrak{M}}}{2r^{6}}$} & $1..384(1)$ & $1.384(1)$\tabularnewline
{\footnotesize $-\frac{\alpha_{1}}{2r^{4}}-\frac{\alpha_{2}-\frac{3\beta_{1}}{\mathfrak{M}}}{2r^{6}}-\frac{\alpha_{112}}{2r^{7}}$} & $1.3840(1)$ & $1.3838(1)$\tabularnewline
\noalign{\vskip2mm}
{\footnotesize $-\frac{\alpha_{1}}{2r^{4}}-\frac{\alpha_{2}-\frac{3\beta_{1}}{\mathfrak{M}}}{2r^{6}}-\frac{\alpha_{112}}{2r^{7}}-\frac{\alpha_{3}-\alpha_{1}\alpha_{2}}{2r^{8}}$} & $1.38397(1)$ & $1.38376(1)$\tabularnewline
\noalign{\vskip2mm}
{\footnotesize $-\frac{\alpha_{1}}{2r^{4}}-\frac{\alpha_{2}-\frac{3\beta_{1}}{\mathfrak{M}}}{2r^{6}}-\frac{\alpha_{112}}{2r^{7}}-\frac{\alpha_{3}-\alpha_{1}\alpha_{2}+\alpha_{1111}}{2r^{8}}$} & $1.3839630(1)$ & $1.3837608(1)$\tabularnewline
\noalign{\vskip2mm}
{\footnotesize $-\frac{\alpha_{1}}{2r^{4}}-\frac{\alpha_{2}^{({\rm no}{\rm \, rel)}}-\frac{3\beta_{1}}{\mathfrak{M}}}{2r^{6}}-\frac{\alpha_{112}}{2r^{7}}-\frac{\alpha_{3}-\alpha_{1}\alpha_{2}+\alpha_{1111}+\frac{72\gamma_{1}}{\mathfrak{M}}}{2r^{8}}$} & $1.383963(1)$ & $1.383761(1)$\tabularnewline
\noalign{\vskip2mm}
{\footnotesize $-\frac{\alpha_{1}}{2r^{4}}-\frac{\alpha_{2}^{({\rm no}{\rm \, QED)}}-\frac{3\beta_{1}}{\mathfrak{M}}}{2r^{6}}-\frac{\alpha_{112}}{2r^{7}}-\frac{\alpha_{3}-\alpha_{1}\alpha_{2}+\alpha_{1111}+\frac{72\gamma_{1}}{\mathfrak{M}}}{2r^{8}}$} & $1.3839631(1)$ & $1.383761(1)$\tabularnewline[2mm]
\hline 
\noalign{\vskip2mm}
$-\frac{\alpha_{1}}{2r^{4}}-\frac{\alpha_{2}-\frac{3\beta_{1}}{\mathfrak{M}}}{2r^{6}}-\frac{\alpha_{112}}{2r^{7}}-\frac{\alpha_{3}-\frac{\beta_{2}}{\mathfrak{M}}-\alpha_{1}\alpha_{2}+\alpha_{1111}+\frac{72\gamma_{1}}{\mathfrak{M}}}{2r^{8}}$ & $1.38396299(1)$ & $1.3837608(1)$\tabularnewline
\end{tabular*}
\end{table*}

\begin{figure}
\caption{Long-range polarization potentials for the ground electronic state
of $^{4}$HeH$^{+}$ in Le~Roy space to emphasize their differences.\label{fig:potentialsInLeRoySpace}}

\includegraphics[width=0.5\textwidth]{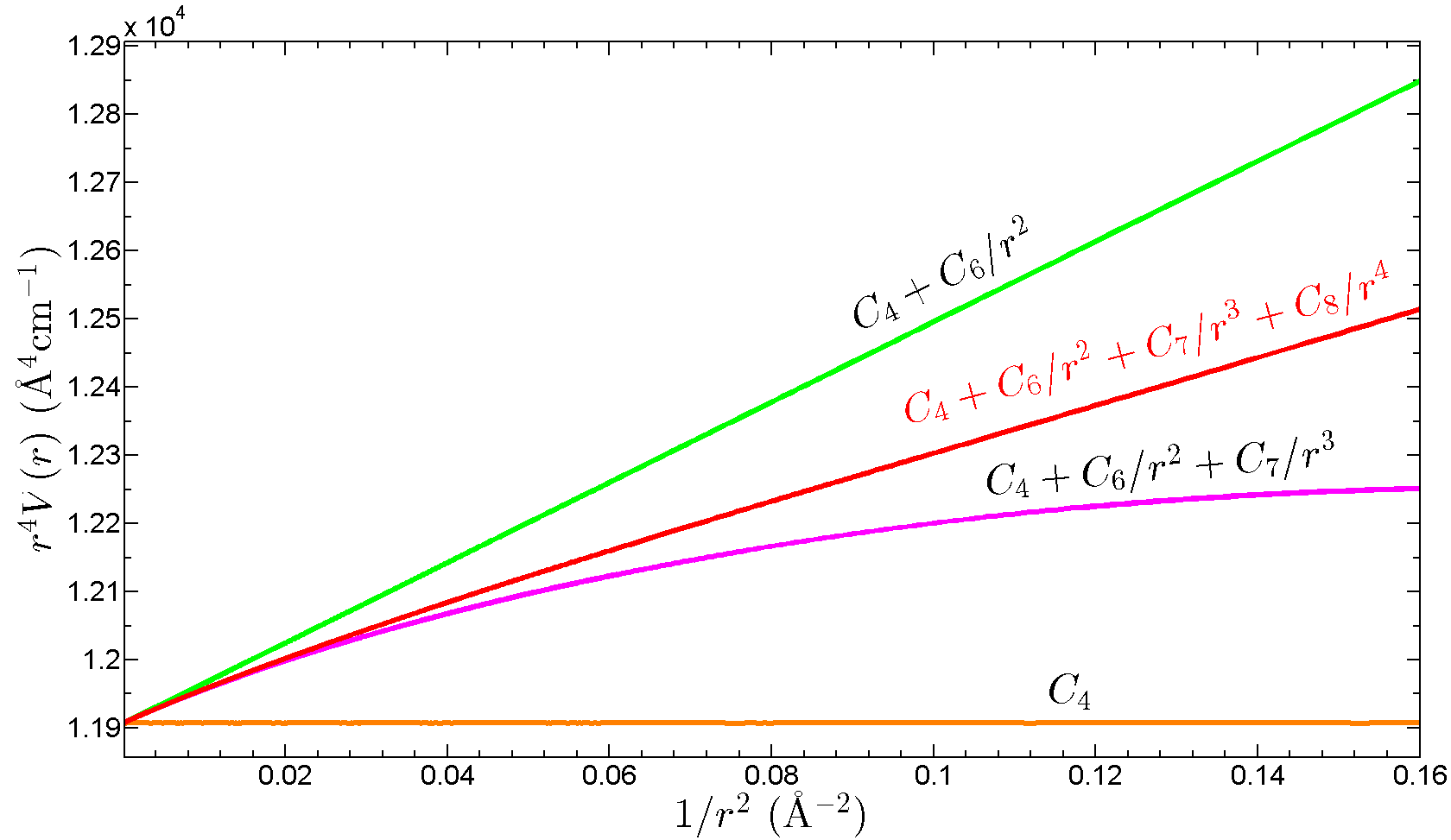}
\end{figure}

Most of the constants appearing in the above polarization function
are not known with very high accuracy, especially for $^{3}$He. We
therefore first calculate these constants with unprecedented precision,
including relativistic ($\alpha_{{\rm FS}}^{2})$ and third-order
QED ($\alpha_{{\rm FS}}^{3}$) effects for $\alpha_{2}$, where $\alpha_{{\rm FS}}$
is the fine structure constant. We then calculate $^{4}$He and $^{3}$He
finite-mass effects, for $\alpha_{2}$, $\alpha_{112}$, $\alpha_{1111}$,
and $\beta_{1}$. The value of $\alpha_{1}$ has been calculated previously
in \cite{Lach2004}, including fourth-order QED ($\alpha_{{\rm FS}}^{4}$)
effects, but finite-mass corrections were only presented for $^{4}$He.
We therefore use this value for $^{4}$He, and calculate the finite-mass
correction for $^{3}$He, and combining these with our calculated
values of the other constants, Eq. \ref{eq:polarizationPotential}
very accurately represents the long-range behavior of the isotopologues
of HeH$^{+}$. 

We then determine how precisely $\alpha_{1}$ can be determined from
a fit to Eq. \ref{eq:polarizationPotential} when varying numbers
of terms are included, and with or without the inclusion of relativistic
and/or QED effects in the adiabatic quadrupole polarizability $\alpha_{2}$.
We find that when relativistic ($\alpha_{{\rm FS}}^{2}$) and/or $\alpha_{{\rm FS}}^{3}$
QED effects are neglected, $\alpha_{1}$ can only be predicted correctly
to at best 6 digits. When these effects are included in the model,
$\alpha_{1}$ can be determined with 8 digits of precision, which
is one order of magnitude more precise than the experimental value
determined in \cite{Schmidt2007}.

The various polarizabilities calculated in this work were determined
following the ideas of numerical calculations presented in \cite{Puchalski2011a}.
The numerical calculation involves a non-relativistic wavefunction
determined variationally. Our wavefunction, consists of explicitly
correlated exponential functions \cite{Korobov2000,*Korobov2002,Puchalski2011a}.
We then calculate the adiabatic and non-adiabatic polarizabilities
as introduced in the optical-potential analysis for Rydberg atoms,
e.g. \cite{Drachman1982,Snow2007}. It is noted that Drachman's definitions
in \cite{Drachman1982} are introduced with Rydberg atomic units,
and here we always use Hartree atomic units.

Since we are using very high-precision and fast calculations with
explicitly correlated exponential functions for the helium atom, the
finite mass corrections are determined here as a difference between
the non-relativistic values obtained in the infinite mass limit, and
those using the relavent nuclear mass. Uncertainties presented for
the leading order non-relativistic contributions as well as for the
non-relativistic finite mass corrections are estimated based on numerical
convergence. The non-relativistic finite mass effects for the $1/r^{8}$
coefficients $\alpha_{3}$, $\beta_{2}$, $\gamma_{1}$ and $\alpha_{1111}$,
and the relativistic finite mass effects for \emph{all} terms, are
expected to be negligible in our application of the long-range potential.
An attempt at the finite mass corrections for $\beta_{2}$ and $\gamma_{1}$
was presented in \cite{Bhatia1994}. However, there are additional
contributions that are expected to be important, e.g. from the finite
mass kinetic energy correction \cite{Pachucki2009}, which shows the
limitations of a simple generalization from the Rydberg states theory,
and possible problems that need to be addressed in a more detailed
theory. 

Since we consider the internuclear potential in the long-range limit,
coefficients of $1/r^{n}$ with lower values of $n$ have to be evaluated
much more accurately, including subtle effects. After $\alpha_{1}$,
the most important contribution to the internuclear potential comes
from the quadruple polarizability $\alpha_{2}$. The non-relativistic
value of $\alpha_{2}$ at leading order has been calculated before
\cite{Bhatia1994,Yan1996}. Here, we calculate the non-relativistic
value with greater precision, and additionally we calculated finite-mass
and relativistic correcions with an analogous approach to the one
used for the dipole polarizability $\alpha_{1}$ in paper \cite{Pachucki2000}.
It is based on perturbation theory using the Breit-Pauli operator
\cite{Bethe1957}. Finally we also included the leading QED effects
as was done for $\alpha_{1}$ in \cite{Pachucki2000}. The only estimation
comes from assuming the Bethe logarithm $\ln k_{0}$ as a known consant
in the QED Hamiltionian. The neglected effects are very demanding
in numerical calculations. We estimated them using 10\% uncertainty
in the $\ln k_{0}$ value. This estimate is embodied in the uncertainty
of the QED correction value in Table \ref{tab:PolarizabilitiesInfiniteMass}.

We have included these calculated values in our compiled listing of
the current most precise values for various polarizabilities (neglecting
finite mass effects) of He in Table \ref{tab:PolarizabilitiesInfiniteMass}.
Finite mass effects are then added to the quantities in Table \ref{tab:PolarizabilitiesInfiniteMass}
in order to obtain the corresponding values for $^{3}$He and $^{4}$He.
Table \ref{tab:polarizabilitiesFiniteMass} shows the size of these
mass polarization corrections for $^{4}$He and $^{3}$He, and the
final polarizabilities for each isotope. Using the values from Table
\ref{tab:polarizabilitiesFiniteMass} and Eq. \ref{eq:polarizationPotential},
we calculate high-precision theoretical long-range potentials for
the ground electronic states of $^{3}$HeH$^{+}$ and $^{4}$HeH$^{+}$. 

In Fig. \ref{fig:potentials}, we compare the recent empirical potential
for $^{4}$HeH$^{+}$ from \cite{Welsh2014}, to the long-range potentials
corresponding to various approximations constructed by truncating
the number of terms used from Eq. \ref{eq:polarizationPotential}.
It is clear in this figure that the highest bound vibrational levels
are in a region where the molecular potential is indeed the long-range
polarization potential of Eq. \ref{eq:polarizationPotential}. Therefore,
fitting the polarization potential (specifically, the parameter representing
the dipole polarizability $\alpha_{1}$) to high-precision spectroscopic
measurements involving the highest vibrational levels will indeed
give us provide an accurate empirical value for $\alpha_{1}$. Since
in this region all of the long-range potentials look the same, it
may appear as if the number of terms used in the polarization model
is irrelavent. However, since we are interested in using $\alpha_{1}$
for a newer, more precise definition of the Boltzmann constant $k_{B}$,
the precision with which $\alpha_{1}$ is empirically determined is
extremely important, and Fig. \ref{fig:potentialsInLeRoySpace} presents
the various approximations to the long-range potential in Le~Roy
space to emphasize that the potentials are in fact significantly different
until at least 5\textcolor{black}{\footnotesize $\mbox{\AA}$}.

In Table \ref{tab:fittedDipolePolarizability} we show that for both
$^{3}$HeH$^{+}$ and $^{4}$HeH$^{+}$ the dipole polarizability
can be determined in agreement with the high-precision (9-digit) theoretical
values by a fit to the full polarization function. This is at least
one order of magnitude more precise than the best experimental value
determined in \cite{Schmidt2007} (see Table \ref{tab:fittedDipolePolarizability}).
The table also shows that in order for this, relativistic and QED
effects must not be neglected in the calculation of the quadrupole
polarizability $\alpha_{2}$, as otherwise the accuracy of the fitted
$\alpha_{1}$ becomes significantly worse. This result is useful for
the study of other atoms beyond He, since it provides an idea of the
size of the effects necessary to include in the long-range potential
in order to fit $\alpha_{1}$ with a certain level of desired precision.
Finally, the conclusion that, with the right level of accuracy in
the long-range potential model, $\alpha_{1}$ can be obtained empirically
with 9 digits of precision, provides testiment to the case for using
the $\alpha_{1}$ of He for the new SI definition of temperature.
\medskip{}

\emph{Acknowledgements. }We gratefully thank Jim Mitroy of Charles
Darwin University (Australia) for his advice, and for being so helpful
even until his final days. We would also like to thank Krzystof Pachucki
of University of Warsaw (Poland), Z-C. Yan of University of New Brunswick
(Canada), Stephen Lundeen of Colorado State University (USA), Richard
J. Drachman and Anand K. Bhatia from NASA (USA), Robert J. Le~Roy
of University of Waterloo (Canada), and Grzegorz Lach from IIMCB (Poland)
for helpful discussions. 

\noindent \begin{center}
{\scriptsize }
\par\end{center}{\scriptsize \par}

\bibliographystyle{apsrev4-1}

\begin{thebibliography}{21}%
\makeatletter
\providecommand \@ifxundefined [1]{%
 \@ifx{#1\undefined}
}%
\providecommand \@ifnum [1]{%
 \ifnum #1\expandafter \@firstoftwo
 \else \expandafter \@secondoftwo
 \fi
}%
\providecommand \@ifx [1]{%
 \ifx #1\expandafter \@firstoftwo
 \else \expandafter \@secondoftwo
 \fi
}%
\providecommand \natexlab [1]{#1}%
\providecommand \enquote  [1]{``#1''}%
\providecommand \bibnamefont  [1]{#1}%
\providecommand \bibfnamefont [1]{#1}%
\providecommand \citenamefont [1]{#1}%
\providecommand \href@noop [0]{\@secondoftwo}%
\providecommand \href [0]{\begingroup \@sanitize@url \@href}%
\providecommand \@href[1]{\@@startlink{#1}\@@href}%
\providecommand \@@href[1]{\endgroup#1\@@endlink}%
\providecommand \@sanitize@url [0]{\catcode `\\12\catcode `\$12\catcode
  `\&12\catcode `\#12\catcode `\^12\catcode `\_12\catcode `\%12\relax}%
\providecommand \@@startlink[1]{}%
\providecommand \@@endlink[0]{}%
\providecommand \url  [0]{\begingroup\@sanitize@url \@url }%
\providecommand \@url [1]{\endgroup\@href {#1}{\urlprefix }}%
\providecommand \urlprefix  [0]{URL }%
\providecommand \Eprint [0]{\href }%
\providecommand \doibase [0]{http://dx.doi.org/}%
\providecommand \selectlanguage [0]{\@gobble}%
\providecommand \bibinfo  [0]{\@secondoftwo}%
\providecommand \bibfield  [0]{\@secondoftwo}%
\providecommand \translation [1]{[#1]}%
\providecommand \BibitemOpen [0]{}%
\providecommand \bibitemStop [0]{}%
\providecommand \bibitemNoStop [0]{.\EOS\space}%
\providecommand \EOS [0]{\spacefactor3000\relax}%
\providecommand \BibitemShut  [1]{\csname bibitem#1\endcsname}%
\let\auto@bib@innerbib\@empty
\bibitem [{\citenamefont {Lach}\ \emph {et~al.}(2004)\citenamefont {Lach},
  \citenamefont {Jeziorski},\ and\ \citenamefont {Szalewicz}}]{Lach2004}%
  \BibitemOpen
  \bibfield  {author} {\bibinfo {author} {\bibfnamefont {G.}~\bibnamefont
  {Lach}}, \bibinfo {author} {\bibfnamefont {B.}~\bibnamefont {Jeziorski}}, \
  and\ \bibinfo {author} {\bibfnamefont {K.}~\bibnamefont {Szalewicz}},\ }\href
  {\doibase 10.1103/PhysRevLett.92.233001} {\bibfield  {journal} {\bibinfo
  {journal} {Physical Review Letters}\ }\textbf {\bibinfo {volume} {92}},\
  \bibinfo {pages} {233001} (\bibinfo {year} {2004})}\BibitemShut {NoStop}%
\bibitem [{\citenamefont {Schmidt}\ \emph {et~al.}(2007)\citenamefont
  {Schmidt}, \citenamefont {Gavioso}, \citenamefont {May},\ and\ \citenamefont
  {Moldover}}]{Schmidt2007}%
  \BibitemOpen
  \bibfield  {author} {\bibinfo {author} {\bibfnamefont {J.}~\bibnamefont
  {Schmidt}}, \bibinfo {author} {\bibfnamefont {R.}~\bibnamefont {Gavioso}},
  \bibinfo {author} {\bibfnamefont {E.}~\bibnamefont {May}}, \ and\ \bibinfo
  {author} {\bibfnamefont {M.}~\bibnamefont {Moldover}},\ }\href {\doibase
  10.1103/PhysRevLett.98.254504} {\bibfield  {journal} {\bibinfo  {journal}
  {Physical Review Letters}\ }\textbf {\bibinfo {volume} {98}},\ \bibinfo
  {pages} {254504} (\bibinfo {year} {2007})}\BibitemShut {NoStop}%
\bibitem [{\citenamefont {{Le Roy}}\ \emph {et~al.}(2009)\citenamefont {{Le
  Roy}}, \citenamefont {Dattani}, \citenamefont {Coxon}, \citenamefont {Ross},
  \citenamefont {Crozet},\ and\ \citenamefont {Linton}}]{LeRoy2009}%
  \BibitemOpen
  \bibfield  {author} {\bibinfo {author} {\bibfnamefont {R.~J.}\ \bibnamefont
  {{Le Roy}}}, \bibinfo {author} {\bibfnamefont {N.~S.}\ \bibnamefont
  {Dattani}}, \bibinfo {author} {\bibfnamefont {J.~A.}\ \bibnamefont {Coxon}},
  \bibinfo {author} {\bibfnamefont {A.~J.}\ \bibnamefont {Ross}}, \bibinfo
  {author} {\bibfnamefont {P.}~\bibnamefont {Crozet}}, \ and\ \bibinfo {author}
  {\bibfnamefont {C.}~\bibnamefont {Linton}},\ }\href {\doibase
  10.1063/1.3264688} {\bibfield  {journal} {\bibinfo  {journal} {The Journal of
  Chemical Physics}\ }\textbf {\bibinfo {volume} {131}},\ \bibinfo {pages}
  {204309} (\bibinfo {year} {2009})}\BibitemShut {NoStop}%
\bibitem [{\citenamefont {Tang}\ \emph {et~al.}(2011)\citenamefont {Tang},
  \citenamefont {Yan}, \citenamefont {Shi},\ and\ \citenamefont
  {Mitroy}}]{Tang2011}%
  \BibitemOpen
  \bibfield  {author} {\bibinfo {author} {\bibfnamefont {L.-Y.}\ \bibnamefont
  {Tang}}, \bibinfo {author} {\bibfnamefont {Z.-C.}\ \bibnamefont {Yan}},
  \bibinfo {author} {\bibfnamefont {T.-Y.}\ \bibnamefont {Shi}}, \ and\
  \bibinfo {author} {\bibfnamefont {J.}~\bibnamefont {Mitroy}},\ }\href
  {\doibase 10.1103/PhysRevA.84.052502} {\bibfield  {journal} {\bibinfo
  {journal} {Physical Review A}\ }\textbf {\bibinfo {volume} {84}} (\bibinfo
  {year} {2011}),\ 10.1103/PhysRevA.84.052502}\BibitemShut {NoStop}%
\bibitem [{\citenamefont {Mayer}\ and\ \citenamefont
  {Mayer}(1933)}]{Mayer1933}%
  \BibitemOpen
  \bibfield  {author} {\bibinfo {author} {\bibfnamefont {J.}~\bibnamefont
  {Mayer}}\ and\ \bibinfo {author} {\bibfnamefont {M.}~\bibnamefont {Mayer}},\
  }\href {\doibase 10.1103/PhysRev.43.605} {\bibfield  {journal} {\bibinfo
  {journal} {Physical Review}\ }\textbf {\bibinfo {volume} {43}},\ \bibinfo
  {pages} {605} (\bibinfo {year} {1933})}\BibitemShut {NoStop}%
\bibitem [{\citenamefont {Dalgarno}\ \emph {et~al.}(1968)\citenamefont
  {Dalgarno}, \citenamefont {Drake},\ and\ \citenamefont
  {Victor}}]{Dalgarno1968}%
  \BibitemOpen
  \bibfield  {author} {\bibinfo {author} {\bibfnamefont {A.}~\bibnamefont
  {Dalgarno}}, \bibinfo {author} {\bibfnamefont {G.}~\bibnamefont {Drake}}, \
  and\ \bibinfo {author} {\bibfnamefont {G.}~\bibnamefont {Victor}},\ }\href
  {\doibase 10.1103/PhysRev.176.194} {\bibfield  {journal} {\bibinfo  {journal}
  {Physical Review}\ }\textbf {\bibinfo {volume} {176}},\ \bibinfo {pages}
  {194} (\bibinfo {year} {1968})}\BibitemShut {NoStop}%
\bibitem [{\citenamefont {Drachman}(1982)}]{Drachman1982}%
  \BibitemOpen
  \bibfield  {author} {\bibinfo {author} {\bibfnamefont {R.}~\bibnamefont
  {Drachman}},\ }\href {\doibase 10.1103/PhysRevA.26.1228} {\bibfield
  {journal} {\bibinfo  {journal} {Physical Review A}\ }\textbf {\bibinfo
  {volume} {26}},\ \bibinfo {pages} {1228} (\bibinfo {year}
  {1982})}\BibitemShut {NoStop}%
\bibitem [{\citenamefont {Woods}\ and\ \citenamefont
  {Lundeen}(2012)}]{Woods2012}%
  \BibitemOpen
  \bibfield  {author} {\bibinfo {author} {\bibfnamefont {S.~L.}\ \bibnamefont
  {Woods}}\ and\ \bibinfo {author} {\bibfnamefont {S.~R.}\ \bibnamefont
  {Lundeen}},\ }\href {\doibase 10.1103/PhysRevA.85.042505} {\bibfield
  {journal} {\bibinfo  {journal} {Physical Review A}\ }\textbf {\bibinfo
  {volume} {85}},\ \bibinfo {pages} {042505} (\bibinfo {year}
  {2012})}\BibitemShut {NoStop}%
\bibitem [{\citenamefont {Tang}\ \emph {et~al.}(2010)\citenamefont {Tang},
  \citenamefont {Zhang}, \citenamefont {Yan}, \citenamefont {Shi},\ and\
  \citenamefont {Mitroy}}]{Tang2010b}%
  \BibitemOpen
  \bibfield  {author} {\bibinfo {author} {\bibfnamefont {L.-Y.}\ \bibnamefont
  {Tang}}, \bibinfo {author} {\bibfnamefont {J.-Y.}\ \bibnamefont {Zhang}},
  \bibinfo {author} {\bibfnamefont {Z.-C.}\ \bibnamefont {Yan}}, \bibinfo
  {author} {\bibfnamefont {T.-Y.}\ \bibnamefont {Shi}}, \ and\ \bibinfo
  {author} {\bibfnamefont {J.}~\bibnamefont {Mitroy}},\ }\href
  {http://www.ncbi.nlm.nih.gov/pubmed/20849171} {\bibfield  {journal} {\bibinfo
   {journal} {The Journal of Chemical Physics}\ }\textbf {\bibinfo {volume}
  {133}},\ \bibinfo {pages} {104306} (\bibinfo {year} {2010})}\BibitemShut
  {NoStop}%
\bibitem [{\citenamefont {Bhatia}\ and\ \citenamefont
  {Drachman}(1994)}]{Bhatia1994}%
  \BibitemOpen
  \bibfield  {author} {\bibinfo {author} {\bibfnamefont {A.~K.}\ \bibnamefont
  {Bhatia}}\ and\ \bibinfo {author} {\bibfnamefont {R.~J.}\ \bibnamefont
  {Drachman}},\ }\href {\doibase 10.1088/0953-4075/27/7/005} {\bibfield
  {journal} {\bibinfo  {journal} {Journal of Physics B: Atomic, Molecular and
  Optical Physics}\ }\textbf {\bibinfo {volume} {27}},\ \bibinfo {pages} {1299}
  (\bibinfo {year} {1994})}\BibitemShut {NoStop}%
\bibitem [{\citenamefont {Dattani}\ and\ \citenamefont {{Le
  Roy}}(2011)}]{Dattani2011}%
  \BibitemOpen
  \bibfield  {author} {\bibinfo {author} {\bibfnamefont {N.~S.}\ \bibnamefont
  {Dattani}}\ and\ \bibinfo {author} {\bibfnamefont {R.~J.}\ \bibnamefont {{Le
  Roy}}},\ }\href {\doibase 10.1016/j.jms.2011.03.030} {\bibfield  {journal}
  {\bibinfo  {journal} {Journal of Molecular Spectroscopy}\ }\textbf {\bibinfo
  {volume} {268}},\ \bibinfo {pages} {199} (\bibinfo {year}
  {2011})}\BibitemShut {NoStop}%
\bibitem [{\citenamefont {Yan}\ \emph {et~al.}(1996)\citenamefont {Yan},
  \citenamefont {Babb},\ and\ \citenamefont {Dalgarno}}]{Yan1996}%
  \BibitemOpen
  \bibfield  {author} {\bibinfo {author} {\bibfnamefont {Z.-C.}\ \bibnamefont
  {Yan}}, \bibinfo {author} {\bibfnamefont {J.~F.}\ \bibnamefont {Babb}}, \
  and\ \bibinfo {author} {\bibfnamefont {A.}~\bibnamefont {Dalgarno}},\ }\href
  {\doibase 10.1103/PhysRevA.54.2824} {\bibfield  {journal} {\bibinfo
  {journal} {Physical Review A}\ }\textbf {\bibinfo {volume} {54}},\ \bibinfo
  {pages} {2824} (\bibinfo {year} {1996})}\BibitemShut {NoStop}%
\bibitem [{\citenamefont {Kar}\ and\ \citenamefont {Ho}(2010)}]{Kar2010}%
  \BibitemOpen
  \bibfield  {author} {\bibinfo {author} {\bibfnamefont {S.}~\bibnamefont
  {Kar}}\ and\ \bibinfo {author} {\bibfnamefont {Y.~K.}\ \bibnamefont {Ho}},\
  }\href {\doibase 10.1103/PhysRevA.81.062506} {\bibfield  {journal} {\bibinfo
  {journal} {Physical Review A}\ }\textbf {\bibinfo {volume} {81}},\ \bibinfo
  {pages} {062506} (\bibinfo {year} {2010})}\BibitemShut {NoStop}%
\bibitem [{\citenamefont {Puchalski}\ \emph {et~al.}(2011)\citenamefont
  {Puchalski}, \citenamefont {Jentschura},\ and\ \citenamefont
  {Mohr}}]{Puchalski2011a}%
  \BibitemOpen
  \bibfield  {author} {\bibinfo {author} {\bibfnamefont {M.}~\bibnamefont
  {Puchalski}}, \bibinfo {author} {\bibfnamefont {U.~D.}\ \bibnamefont
  {Jentschura}}, \ and\ \bibinfo {author} {\bibfnamefont {P.~J.}\ \bibnamefont
  {Mohr}},\ }\href {\doibase 10.1103/PhysRevA.83.042508} {\bibfield  {journal}
  {\bibinfo  {journal} {Physical Review A}\ }\textbf {\bibinfo {volume} {83}},\
  \bibinfo {pages} {042508} (\bibinfo {year} {2011})}\BibitemShut {NoStop}%
\bibitem [{\citenamefont {Korobov}(2000)}]{Korobov2000}%
  \BibitemOpen
  \bibfield  {author} {\bibinfo {author} {\bibfnamefont {V.}~\bibnamefont
  {Korobov}},\ }\href {\doibase 10.1103/PhysRevA.61.064503} {\bibfield
  {journal} {\bibinfo  {journal} {Physical Review A}\ }\textbf {\bibinfo
  {volume} {61}},\ \bibinfo {pages} {064503} (\bibinfo {year}
  {2000})}\BibitemShut {NoStop}%
\bibitem [{\citenamefont {Korobov}(2002)}]{Korobov2002}%
  \BibitemOpen
  \bibfield  {author} {\bibinfo {author} {\bibfnamefont {V.~I.}\ \bibnamefont
  {Korobov}},\ }\href {\doibase 10.1103/PhysRevA.66.024501} {\bibfield
  {journal} {\bibinfo  {journal} {Physical Review A}\ }\textbf {\bibinfo
  {volume} {66}},\ \bibinfo {pages} {024501} (\bibinfo {year}
  {2002})}\BibitemShut {NoStop}%
\bibitem [{\citenamefont {Snow}\ and\ \citenamefont
  {Lundeen}(2007)}]{Snow2007}%
  \BibitemOpen
  \bibfield  {author} {\bibinfo {author} {\bibfnamefont {E.~L.}\ \bibnamefont
  {Snow}}\ and\ \bibinfo {author} {\bibfnamefont {S.~R.}\ \bibnamefont
  {Lundeen}},\ }\href {\doibase 10.1103/PhysRevA.76.052505} {\bibfield
  {journal} {\bibinfo  {journal} {Physical Review A}\ }\textbf {\bibinfo
  {volume} {76}},\ \bibinfo {pages} {052505} (\bibinfo {year}
  {2007})}\BibitemShut {NoStop}%
\bibitem [{\citenamefont {Pachucki}\ and\ \citenamefont
  {Komosa}(2009)}]{Pachucki2009}%
  \BibitemOpen
  \bibfield  {author} {\bibinfo {author} {\bibfnamefont {K.}~\bibnamefont
  {Pachucki}}\ and\ \bibinfo {author} {\bibfnamefont {J.}~\bibnamefont
  {Komosa}},\ }\href@noop {} {\bibfield  {journal} {\bibinfo  {journal}
  {Journal of Chemical Physics2}\ }\textbf {\bibinfo {volume} {130}} (\bibinfo
  {year} {2009})}\BibitemShut {NoStop}%
\bibitem [{\citenamefont {Pachucki}\ and\ \citenamefont
  {Sapirstein}(2000)}]{Pachucki2000}%
  \BibitemOpen
  \bibfield  {author} {\bibinfo {author} {\bibfnamefont {K.}~\bibnamefont
  {Pachucki}}\ and\ \bibinfo {author} {\bibfnamefont {J.}~\bibnamefont
  {Sapirstein}},\ }\href {\doibase 10.1103/PhysRevA.63.012504} {\bibfield
  {journal} {\bibinfo  {journal} {Physical Review A}\ }\textbf {\bibinfo
  {volume} {63}},\ \bibinfo {pages} {012504} (\bibinfo {year}
  {2000})}\BibitemShut {NoStop}%
\bibitem [{\citenamefont {Bethe}\ and\ \citenamefont
  {Salpeter}(1957)}]{Bethe1957}%
  \BibitemOpen
  \bibfield  {author} {\bibinfo {author} {\bibfnamefont {H.~A.}\ \bibnamefont
  {Bethe}}\ and\ \bibinfo {author} {\bibfnamefont {E.~E.}\ \bibnamefont
  {Salpeter}},\ }\href@noop {} {\emph {\bibinfo {title} {{Quantum mechanics of
  one- and two-electron atoms, Ch. 39}}}}\ (\bibinfo  {publisher} {Springer,
  Berlin},\ \bibinfo {year} {1957})\BibitemShut {NoStop}%
\bibitem [{\citenamefont {Welsh}\ \emph {et~al.}(2014)\citenamefont {Welsh},
  \citenamefont {Puchalski}, \citenamefont {Lach}, \citenamefont {Tung},
  \citenamefont {Adamowicz},\ and\ \citenamefont {Dattani}}]{Welsh2014}%
  \BibitemOpen
  \bibfield  {author} {\bibinfo {author} {\bibfnamefont {S.}~\bibnamefont
  {Welsh}}, \bibinfo {author} {\bibfnamefont {M.}~\bibnamefont {Puchalski}},
  \bibinfo {author} {\bibfnamefont {G.}~\bibnamefont {Lach}}, \bibinfo {author}
  {\bibfnamefont {W.~C.}\ \bibnamefont {Tung}}, \bibinfo {author}
  {\bibfnamefont {L.}~\bibnamefont {Adamowicz}}, \ and\ \bibinfo {author}
  {\bibfnamefont {N.~S.}\ \bibnamefont {Dattani}},\ }in\ \href@noop {} {\emph
  {\bibinfo {booktitle} {Proceedings of the International Symposium on
  Molecular Spectroscopy}}}\ (\bibinfo {year} {2014})\ p.\ \bibinfo {pages}
  {FA02}\BibitemShut {NoStop}%
\end{thebibliography}

\end{document}